\documentclass[final,english]{bullsrsl}[2024/09/30]
\RequirePackage{hyperref}
\RequirePackage{lineno}
\usepackage[latin1]{inputenc}
\usepackage[T1]{fontenc}
\usepackage{natbib} % Add "numbers" option if natbib is used
\usepackage{graphicx}

\usepackage{color}
\definecolor{new}{rgb}{0.8,0.2,0.2}

\definecolor{todo}{rgb}{0.8,0.2,0.3}

\definecolor{comm}{rgb}{0,0.7,0}

\definecolor{idea}{rgb}{0,0.5,1}

\definecolor{old}{rgb}{1,0.5,0.5}

\begin{document}
\title{BRITE Nascent Binaries}
\author{Andrzej}{Pigulski}
\affiliation{Instytut Astronomiczny, Uniwersytet Wroc{\l}awski, Kopernika 11, 51-622 Wroc{\l}aw, Poland}
\correspondance{andrzej.pigulski@uwr.edu.pl}
\date{21st September 2024}
\maketitle

% Abstract of the paper in the same language as the paper
\begin{abstract}
Nascent binaries (NBs) are binary systems with very low mass ratios, less than $\sim$0.2, in which the more massive component is an O- or B-type main-sequence star, while the secondary is a star contracting onto the main sequence. NBs are of interest because they can help to understand the formation of small-mass ratio systems and shed light on the origin of low-mass X-ray binaries, millisecond pulsars and type Ia supernovae. In photometry, short-period NBs show a strong irradiation effect due to the large difference between the effective temperatures of the components and the strong irradiation of a cool secondary by a hot primary. In spectroscopy, they usually appear as single-lined spectroscopic binaries. In the present paper, we summarize the status of our knowledge of Galactic nascent binaries and characterize two new members of this group, c$^2$\,Sco and V390\,Pup, for which photometric data were obtained by the BRIght Target Explorer (BRITE) nano-satellite mission.
\end{abstract}


%\msccodes{65F15 65G50 15-04 15B99}

%\section{Section -- Level 1 title (Times New Roman, bold, 14 pts)}
\section{Introduction}
As a result of their search for small-mass ratio systems among massive eclipsing stars in the Large Magellanic Cloud (LMC), \cite{2015ApJ...801..113M} found a group of 22 stars characterized by narrow eclipse widths and low eclipse depths ratios. Such systems were also characterized by a strong irradiation (reflection) effect and their orbital periods ranged between 3 and 8.5 days. An estimation of the parameters of the components of such systems led these authors to the conclusion that they consist of a massive binary of O or early B spectral type and a cool pre-main sequence (PMS) secondary, strongly irradiated by the hot companion. Because of their young age, the systems were dubbed nascent binaries (NBs). 
In fact, NBs are strongly related to so-called Lindroos systems \citep[][and earlier papers of the series]{1986A&A...156..223L}, Galactic visual binary systems comprised of an early-type main-sequence (MS) star and a late-type secondary. \cite{1986A&A...156..223L} found that in 37 systems the ages of secondaries are shorter than the time of contraction onto the zero-age MS (ZAMS) and described them as post-T~Tauri stars contracting onto the MS. These stars can be therefore regarded as NBs in systems with wide orbits. 
%\comm{[please specify the typical range of orbital periods for these (Lindroos) systems]}

If we adopt that nascent binaries are systems consisting of a massive MS star and a low-mass pre-MS star, we can conclude that such systems can be discovered by several complementary methods. Close systems of this type can be discovered following \cite{2015ApJ...801..113M}, that is, by searching for eclipsing binaries with a hot primary component, narrow eclipses and a strong irradiation effect. In general, however, such systems need not be eclipsing, as we will comment in Sect.~\ref{Galactic-NBs}. 
%\comm{[It is a generalization.. however, it strongly depends on the current mass ratio, moreover, we should not forget about possible post-mass transfer companions. If there is a neutron star or black hole nearby, we will not have eclipses as well (of course it is a very exotic scenario). However, what I noticed in the present version of the proceeding --- there is no discussion regarding the possible overlap of post- and pre-main sequence companions. I think you should be more careful about it, and add a few sentences in the Introduction about this topic.]} 
Spectroscopically, NBs, both eclipsing and non-eclipsing, will be mostly single-lined spectroscopic binaries (SB1). Thus, it seems that a natural method to search for NBs, both eclipsing and non-eclipsing, is to analyse the photometry of SB1 systems, a method we applied in Sect.\,\ref{BRITE-NBs} to observations from BRITE satellites. Finally, such systems can be found among relatively nearby visual systems for which the secondary components can be characterized observationally, i.e.~in the way \cite{1986A&A...156..223L} did. One of the key observations that can in such systems indicate the PMS status of the secondary component is a detection of an X-ray emission of the secondary, like in the $\beta$~Cru system \citep{2008MNRAS.386.1855C}, a triple system containing a PMS star. % The PMS star is $\beta$~Cru D.

\section{Galactic nascent binaries}\label{Galactic-NBs}
Given the relatively small number of nearby B-type MS stars \cite{2015ApJ...801..113M} concluded that ``{\it It is therefore not surprising that we have not yet observed in the Milky Way the precise counterparts to our reflecting eclipsing binaries with B-type MS primaries and low-mass pre-MS companions.}'' Galactic NBs are known, however, including Lindroos stars mentioned above. One of the first discovered Galactic NB systems was 16 (EN) Lac, an SB1 system composed of the B2\,IV primary component, which pulsates as a $\beta$~Cephei-type star and a low-mass component. It is an eclipsing binary with the orbital period of 12.097~d. Modelling of the primary eclipse and other considerations led \cite{1988AcA....38..401P} to conclude that the secondary is a likely PMS star, which was later confirmed by \cite{2015MNRAS.454..724J}. The system is too wide to show significant irradiation effect, however. A very shallow secondary eclipse was found recently in this system with the precise Transiting Exoplanet Survey Satellite (TESS) data \citep{2022MNRAS.513.3191S}. The next Galactic NB is the bright eclipsing binary $\lambda$~Sco, in which the primary pulsates as a $\beta$~Cep-type star. The secondary component of this triple star was found to be a PMS star by \cite{2004A+A...427..581U} and \cite{2006MNRAS.370..884T}.

The increasing precision of space-based data allowed the discovery of further Galactic NBs. Another eclipsing NB was $\mu$~Eri, consisting of a B5\,IV primary pulsating in $g$-modes (an SPB-type variability) and a low-mass secondary \citep{2013MNRAS.432.1032J}. Using the Microvariability and Oscillations of STars (MOST) satellite data, these authors concluded that the secondary can be a PMS star. Two bright non-eclipsing NBs, $\nu$~Centauri and $\gamma$~Lupi, were discovered using space-based BRIght Target Explorer (BRITE) data \citep{2021MNRAS.503.5554J}. In both stars, two-colour BRITE data revealed photometric variability with the same period as the variability of their radial velocities. This variability was attributed to the irradiation effect as the period of variability was consistent with the orbital period and the red-filter amplitude was higher than the blue-filter amplitude, as expected for such systems. This showed that non-eclipsing NBs can be discovered provided that radial-velocity data are available. This type of variability is designated `R' in the Variability Star Index \citep[VSX,][]{2006SASS...25...47W}.
%\comm{[Is this information essential?]}

The TESS data have proven to be an excellent source of precision photometry for discovering Galactic NBs, in particular those similar to \cite{2015ApJ...801..113M} stars, i.e.~with strong irradiation effect. Such stars have been found by \cite{2021ApJ...910..133S} and more recently by \cite{2023MNRAS.525.1641N}. It is also easy to identify such candidates among the stars presented by \cite{2024ApJS..272...25E}, although in each of these cases modelling and, in particular, an estimate of the age of these systems is needed to be able to verify whether the secondary components are indeed PMS stars. Radial velocities are also a crucial element: indeed, only the combined analysis of photometry and velocities can confirm that the O- or B-type star is the hottest component of the system and exclude post-interaction cases with a hot stripped star paired to the O- or B-type star.

\section{New nascent binaries with BRITE observations}\label{BRITE-NBs}
In this study, we have chosen to use BRITE Constellation data \citep{2014PASP..126..573W,2016PASP..128l5001P,2017A&A...605A..26P} as the main data source. Currently, such data are available for 716 bright stars \citep{2024A&A...683A..49Z}. Bright stars tend to have observations of many kinds, including spectroscopic ones, which allows us to more easily determine the parameters of the components and, in particular, to verify whether the secondary components are PMS stars. Using the 9th Catalogue of Spectroscopic Binary Orbits \citep{2004A&A...424..727P} we checked which of the stars observed by BRITEs are known to be SB1 systems. The search resulted in the selection of 39 SB1 systems with primaries of spectral type B5 or earlier. Of these stars, 16 had orbital periods shorter than 15~d, giving a chance of detecting an irradiation effect. The sample included five known Galactic NBs mentioned in Sect.\,\ref{Galactic-NBs}, 16 (EN) Lac, $\mu$~Eri, $\nu$~Cen, $\gamma$~Lup~A and $\lambda$~Sco. We carefully checked the BRITE and TESS photometry of the remaining 11 stars and found two candidates for Galactic nascent binaries. These are c$^2$\,Sco (Sect.\,\ref{c2Sco}) and V390\,Pup (Sect.\,\ref{v390pup}).

\subsection{HD\,145482 (13 c$^2$\,Sco)}\label{c2Sco}
HD\,145482 (13 c$^2$ Sco, $V=$ 4.6~mag, B2.5\,Vn) was found spectroscopically variable by \cite{1928PLicO..16....1C}, which was later confirmed by \cite{1960MNRAS.121..263B}, \cite{1963ApJ...137..824V} and \cite{1969BAN....20..204V}. The first spectroscopic orbit was derived by \cite{1987ApJS...64..487L}, who concluded that radial velocities of this star vary with a period of 5.780531(58) d.

HD\,145482 was one of the targets of BRITE-Heweliusz red-filter BRITE satellite in Field Sco~I. The observations spanned 63 days in 2015. The data show clearly the variability with a period of about 5.77~d \citep{2024A&A...683A..49Z}, similar to the spectroscopic period. The star was also observed by three other space missions: The Solar Mass Ejection Imager \citep[SMEI,][]{2003SoPh..217..319E} with observations spanning almost 8 years between 2003 and 2010, Kepler K2 mission with observations taken during Campaign 2 in 2014 and spanning 75 days, and TESS mission, which observed the star in Sector 12. The TESS observations spanned 24 days in 2019. First, we have analyzed the data sets separately. All they reveal clear variability with the same period of about 5.77~d. We then normalized all four data sets by their amplitudes and combined. The final fit to the combined SMEI, K2, BRITE and TESS photometric data, which included also the first harmonic of the main frequency, resulted in the period of 5.77101(30)~d. This is in perfect agreement with the period of 5.77102~d derived by Sebastian A.~Otero and reported in the VSX web page and slightly off the period of 5.814(5)~d found by \cite{2016MNRAS.457.3724B} from the K2 data (the star is EPIC\,202909059) and 5.804~d derived by \cite{2022MNRAS.513.1429S}. The photometric data, phased with the derived period, are shown in Fig.~1.
\begin{figure}[!ht]
\label{c2Sco-phased}
\centering
\includegraphics[]{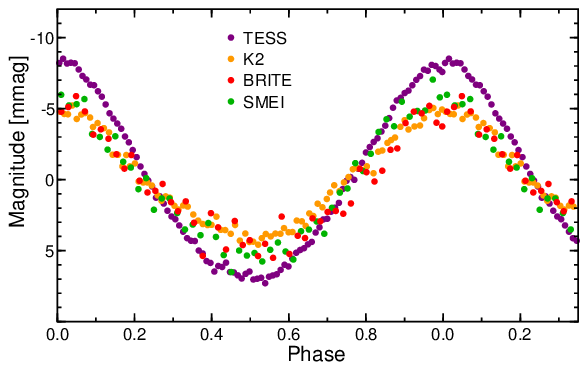}
\begin{minipage}{12cm}
\centering
\caption{BRITE (red dots), Kepler K2 (orange dots) and TESS (violet dots) and SMEI (green dots) data of HD\,145482 freed from the strongest pulsation terms and phased with the period of 5.77101~d. Phase 0.0 corresponds to HJD\,2456458.26, the time of the maximum light of the main term in the combined photometric data.}
\end{minipage}
\end{figure}

The semi-amplitudes of the variability of the main term are equal to  4.38(5), 4.64(11), 5.54(15) and 7.463(7)\,mmag for the K2, BRITE, SMEI and TESS observations, respectively. This sequence of increasing amplitudes corresponds to the increasing median wavelengths of the corresponding passbands. The amplitude is the largest for the TESS data, as the TESS passband has the longest median wavelength of about 800~nm. This behaviour (larger amplitudes at longer wavelengths) is the same as for $\nu$~Cen and $\gamma$~Lup~A \citep{2021MNRAS.503.5554J}, which is a good argument in favour of the explanation of the variability seen in Fig.~1 by the irradiation effect.
\begin{figure}[!ht]
\centering
\includegraphics[]{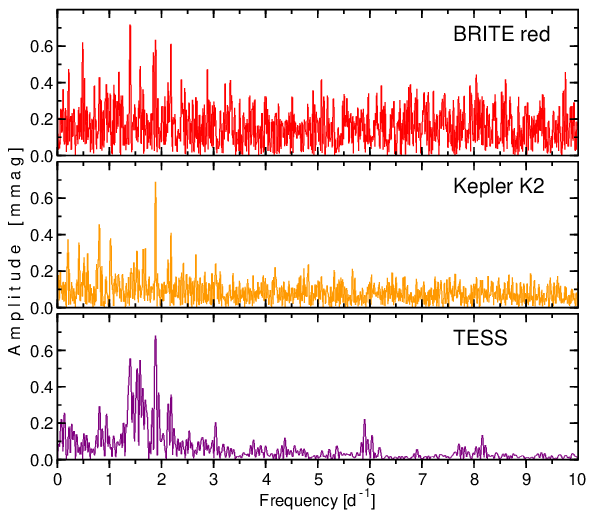}
\begin{minipage}{12cm}
\centering
\caption{Fourier frequency spectra of the residuals after removing the variability with the period of 5.77~d for BRITE (top), Kepler K2 (middle) and TESS (bottom) data.}
\end{minipage}
\end{figure}

In addition to this nearly sinusoidal variability with the period of about 5.77~d, the residual frequency spectra show additional variability, mainly in the low-frequency region (Fig.~2). In particular, the strongest term that can be seen in all three data sets shown in Fig.~2 (the SMEI data are too scattered to reveal this variability) has the frequency of 1.8855(6) in the Kepler data and 1.8845(2) in the TESS data. A peak at this frequency can be also seen in the BRITE data, although the BRITE data are more scattered than the TESS and K2 data. These frequencies correspond to the period of about 0.5305~d, which is in agreement with the period of 0.53042~d reported in VSX. It can be seen from Fig.\,2, however, that there are more frequencies that can be extracted from these data. Given the values of these frequencies and the spectral type of the primary, it is very likely that these frequencies represent $g$-modes excited in the primary. It is also possible that some high-frequency peaks, seen around 6 and 8~d$^{-1}$ in the TESS data in Fig.~2 represent $p$-modes. It is therefore quite likely that the primary in this system is an SPB-type variable, maybe even a hybrid SPB/$\beta$~Cep-type variable.

The spectroscopic data of HD\,145482 are scattered and most of them was obtained more than 45 years ago. They consist of three spectra made in 1915 and 1916 at Lick Observatory, USA \citep{1928PLicO..16....1C}, seven spectra made in 1955 at McDonald Observatory, USA \citep{1963ApJ...137..824V}, three spectra made in 1956 and 1957 at Mount Stromlo Observatory, Australia \citep{1960MNRAS.121..263B}, 13 spectra obtained by \cite{1969BAN....20..204V} in 1966 at Radcliffe Observatory, South Africa, and eight spectra taken in the years 1974\,--\,1977 at Cerro Tololo Interamerican Observatory (CTIO), Chile, and Kitt Peak National Observatory, USA \citep{1987ApJS...64..487L} and a single spectrum obtained with UVES at VLT in Paranal Observatory, Chile \citep{2023ApJS..266...11B}. Three spectra were also obtained with the CHIRON spectrograph at CTIO \citep{2016AJ....151....3G}, but radial velocities of the primary were not given. We transformed all times of observations to Heliocentric Julian Day (HJD) and phased the reported radial velocities with the same period and initial epoch as the photometric data shown in Fig.~1. The resulting radial-velocity curve is shown in Fig.~3. It can be seen from this figure that, despite that some radial-velocity data are discrepant, the maximum of the radial velocity curve is around phase 0.25. Since phase 0.0 corresponds to the light maximum (Fig.~1), this is exactly the relationship between the phases of the light variability and the variation in radial velocities that can be expected when the source of photometric variation is the irradiation effect. 
\begin{figure}[!ht]
\centering
\includegraphics[]{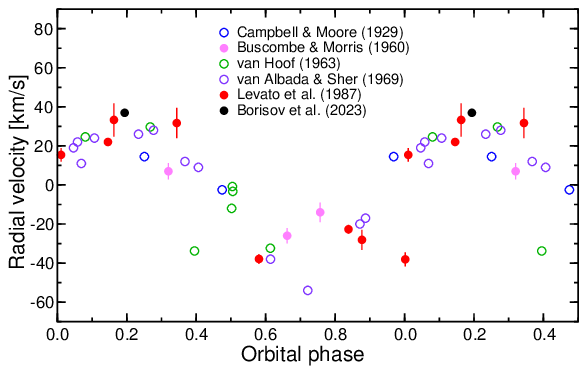}
\begin{minipage}{12cm}
\centering
\caption{Radial-velocities of the primary component of c$^2$\,Sco phased with the orbital period of 5.77101~d. Phase 0.0 corresponds to the same time as in Fig.~1.}
\end{minipage}
\end{figure}

An additional confirmation of c$^2$\,Sco as a new NB comes from the location of the secondary in the mass\,--\,age diagram, shown in Fig.~8. The parameters, secondary's mass of 1.1$^{+0.07}_{-0.08}$\,M$_\odot$ 
%\comm{[number of significant digits differ between value and error.]} 
and age of 8$^{+6}_{-3}$\,Myr, were taken from \cite{2016AJ....152...40G}. It is clear from this plot 
%\comm{[See the comment below, is this plot so universal that if we place there the stars we can understand everything?]} 
that the secondary is a PMS star. Therefore, we conclude that, in addition to $\nu$~Cen and $\gamma$~Lup~A, c$^2$\,Sco is the third known non-eclipsing Galactic nascent binary with its photometric variability dominated the the irradiation effect.

\subsection{HD\,62747 (V390~Pup)}\label{v390pup}
HD\,62747 ($V=$ 5.6~mag, B1.5\,III) was first searched for variability by \cite{1977AcA....27..365J}, but the result was inconclusive due to the variability of one of the comparison stars. It was found an eclipsing variable from the Hipparcos data with a period of 3.9279~d \citep{1997ESASP1200.....E}. The star was observed with red-filter BRITE-Toronto satellite in the Field CMa~I. The observations spanned 105~days between December 10, 2015 and March 24, 2016. The BRITE data confirmed the eclipsing nature of the variability \citep{2024A&A...683A..49Z}. Figure 4 shows the Hipparcos and BRITE data phased with the orbital period of 3.9276888~d, derived below. The BRITE data revealed shallow secondary eclipse which was not covered by the Hipparcos data.
\begin{figure}[!ht]
\centering
\includegraphics[]{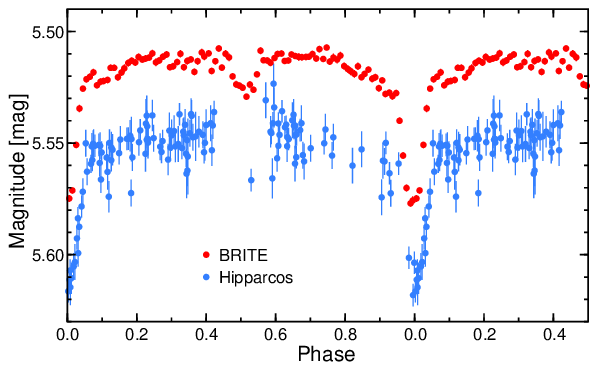}
\begin{minipage}{12cm}
\centering
\caption{Photometry of V390~Pup from Hipparcos (blue dots) and BRITE (red dots) phased with the period of 3.9276888~d. The BRITE data were averaged in 0.01 phase intervals and shifted in magnitude. Phase 0.0 corresponds to HJD\,2448501.04.}
\end{minipage}
\end{figure}

The most abundant precise photometric observations of V390\,Pup were, however, obtained by the TESS satellite, which observed this star in three sectors, Sector 7 (late 2018 -- early 2019), 34 (in 2021) and 61 (in 2023). The orbital period of the star was derived by several investigators: 3.89(3)~d \citep{2020A&A...639A..81B}, 3.92720311~d \citep{2021ApJ...912..123J}, 3.92609~d \citep{2021A&A...652A.120I}, 3.9279~d \citep{2022ApJS..259...50S} and 3.92733(35) \citep{2022ApJS..258...16P}.

We carried out our own analysis of the TESS photometry of V390~Pup and derived the orbital period of 3.9276888(6)~d, which is based on the observations from all three sectors mentioned above, spanning slightly more than four years. We also modelled the light curve using the JKTEBOP modelling code \citep{2012ascl.soft07013S}. The most important parameters derived from this modelling were the following (we assumed that the orbit is circular): inclination of the orbit, $i=73.8(7)^{\rm o}$, the primary's radius, $R_1$, in terms of the relative distance between the stars, $a$,  $r_1 = R_1/a = 0.331(7)$ and the ratio of the radii, $k=R_2/R_1 = 0.245(16)$. The comparison of the TESS light curve with the fitted model is shown in Fig.~5. It is clear that in addition to the significant irradiation effect, the ellipsoidal effect also contributes to the out-of-eclipse variability, which makes the light curve relatively flat in the vicinity of the secondary eclipse.
\begin{figure}[!ht]
\centering
\includegraphics[]{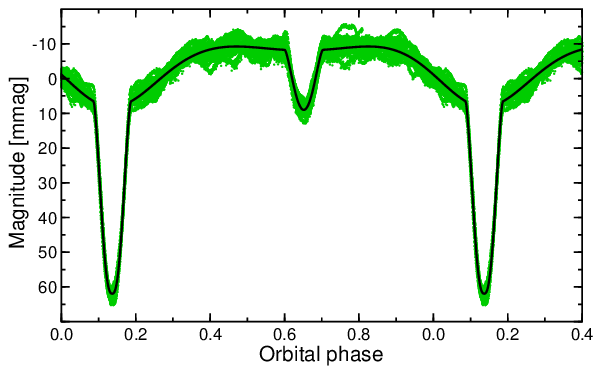}
\begin{minipage}{12cm}
\centering
\caption{TESS light curve of V390~Pup freed from the strongest pulsations and phased with the period of 3.9276888~d. The black curve is the model fitted with the JKTEBOP code.}
\end{minipage}
\end{figure}

\cite{2020A&A...639A..81B} already noted that some low-frequency variability is present out of eclipse in the light curve of V390~Pup. We therefore subtracted the orbital frequency and its harmonics and calculated Fourier frequency spectrum, which is shown in Fig.~6. It can be seen in this figure that the star shows variability which, given the spectral type of the primary, can be attributed to both $g$-modes (lower frequencies) and $p$-modes (higher frequencies). In addition, there is a clear increase of power towards the lowest frequencies, an indication of stochastic variability observed in many early B-type stars \citep[e.g.][]{2019A&A...621A.135B}.
\begin{figure}[!ht]
\centering
\includegraphics[]{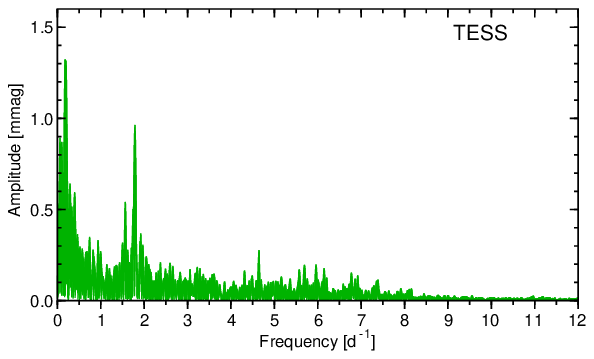}
\begin{minipage}{12cm}
\centering
\caption{Fourier frequency spectrum of the TESS light curve of V390~Pup after removing the orbital frequency and its harmonics.}
\end{minipage}
\end{figure}

Spectroscopy of V390~Pup is scarce and was obtained mainly to derive its spectral type, rotational velocity or other atmospheric parameters \citep{1969ApJ...157..313H,2004MNRAS.351..745L,2006A&A...452..945T,2014A&A...562A.135S}. There is, however, no time-series spectroscopy of this star aimed at detection of the variability of its radial velocities. Nevertheless, the star is an SB1 system since no information on the detection of secondary's lines was announced. In cooperation with Dr.~L.Vanzi and P.~Torres from Pontificia Universidad Cat\'olica (PUC) de Chile, a series of spectroscopic observations of the system was obtained using FIDEOS spectrograph \citep{2018MNRAS.477.5041V}. Full analysis of the combined spectroscopy and photometry of V390~Pup will be published elsewhere. Here, we use only the preliminary value of the half-range of the radial-velocity variations derived from these observations and equal to $K_1 = 34.7$\,$\pm$\,1.9~km\,s$^{-1}$, which is needed to get a rough estimate of the secondary's mass. This value of $K_1$ results in the mass function equal to $f(M) = 0.0170(28)$\,M$_\odot$. In the low-mass ratio eclipsing systems, secondary's mass and radius can be derived quite precisely because of the weak dependence of these parameters on the (assumed) primary's mass. An example of this kind of estimation is shown in Fig.~7.
\begin{figure}[!ht]
\centering
\includegraphics[]{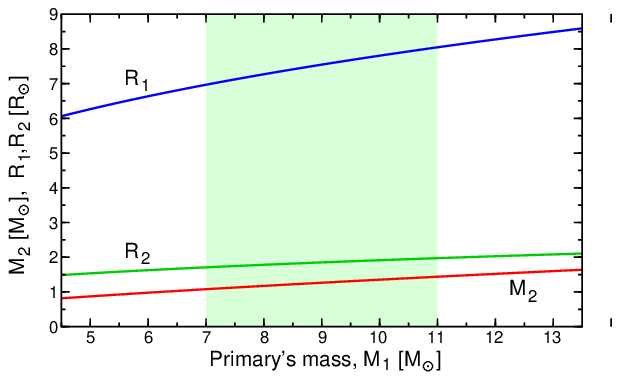}
\begin{minipage}{12cm}
\centering
\caption{Values of the secondary's mass ($M_2$, red line), the secondary's radius ($R_2$, green line) and the primary's radius ($R_1$, blue line) as a function of the primary's mass in the V390~Pup system. The shaded area marks the assumed range of the primary's mass.}
\end{minipage}
\end{figure}

Figure 7 shows the secondary's mass in the V390~Pup system, $M_2$, and the radii of both components, $R_1$ and $R_2$, as a function of the primary's mass. Assuming primary's mass specific for its spectral type and luminosity class, $M_1=9\,\pm\,2$\,M$_\odot$, and using relations shown in Fig.~7, we get $M_2 = 1.27^{+0.17}_{-0.19}$\,M$_\odot$, $R_2=1.85^{+0.12}_{-0.14}$\,R$_\odot$, and $R_1=7.55^{+0.50}_{-0.58}$\,R$_\odot$. An estimated age of the primary, 16.2\,$\pm$1.0~Myr \citep{1978A+AS...32..409G}, places the secondary in the area of PMS stars, as shown in Fig.~8. The secondary's radius is significantly larger than expected for the zero-age main sequence star with secondary's mass, which can also be an indication of its PMS nature. 

\begin{figure}[!ht]
\centering
\includegraphics[]{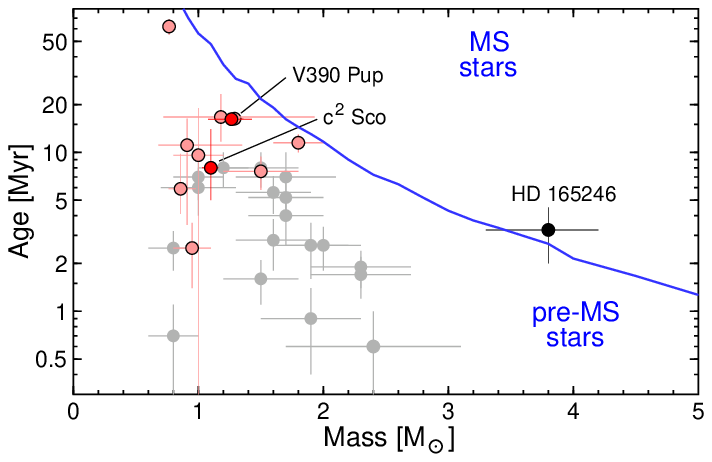}
\begin{minipage}{12cm}
\centering
\caption{Mass\,--\,age diagram for the PMS secondaries in NBs. The grey dots mark the LMC NBs of \cite{2015ApJ...801..113M}. The light-red dots are Galactic NBs from Table 1. Three stars discussed in the text, the two new NBs (red dots) and HD\,165246 (black dot), are labelled. The blue line defines the border between the pre-MS and MS regions. The values of the pre-MS phase duration were derived from the models of \cite{2011A+A...533A.109T} for solar metallicity.
%\comm{[or which metallicities (you plotted Galactic PMS systems plus LMC ones.. is that ok? If so, elaborate please!), initial conditions, and parameters .. So far I don't understand why we should trust this line. Are there any other lines could exist there? :)]}
}
\end{minipage}
\end{figure}

\section{Conclusions}
The presently known sample of 11 Galactic nascent binaries, of which eight are eclipsing and three are non-eclipsing, is summarized in Table 1. In the present paper, we added two members to this group, c$^2$\,Sco, a non-eclipsing NB, and V390~Pup, an eclipsing NB, both observed with BRITE satellites. Table 1 provides for all these stars orbital periods, spectral types of primaries, (SpT)$_1$, the masses of secondaries, $M_2$, mass ratios, $q=M_2/M_1$ and ages.  All known Galactic NBs are also shown in the mass\,--\,age diagram (Fig.~8). In addition to the NBs we also show an example of a star, which, similarly to the short-period NBs, shows a significant irradiation effect. This is HD\,165246 \citep{2021MNRAS.503.1124J}. With the secondary's mass 
%\comm{[Dynamic mass/ evolutionary mass? ]} 
$M_2=3.8_{-0.5}^{+0.4}$\,M$_\odot$ and age estimation between 2 and 4.5~Myr, derived by these authors, the star locates close to the zero-age main sequence and therefore cannot be regarded as a nascent binary. 
%\comm{[I don't get the point of how does position of the primary close to the ZAMS depends on the nature of the binary (NB or not NB). In Naze et al. (2023) all systems are located close to the ZAMS.]} 
This shows that, although the irradiation effect can be used to find candidates for NBs, the final verification of the evolutionary status of the secondary component should involve modelling of the light curve and measurements of radial velocities. 
%\comm{[They are not just very useful, the RV curve is extremely important for understanding if we are dealing with OB+PMS systems or OB+stripped star systems, only phase-folded RV- and light-curves could tell us reflection from which object we observe while we look on the lightcurve! Without the RV curve we simply don't know if we see the reflection of the OB star or we see the reflection of the more hot stripped star on the OB star... (see Section 3.3. in Naze et al. 2023)]}
\begin{table}[!ht]
\centering\footnotesize
\begin{minipage}{88mm}
\caption{Presently known Galactic nascent binaries with orbital periods shorter than 15~days.
%\comm{[Please add the information about the SpT of the primaries and the mass ratio for each system.]}
}
\end{minipage}
\bigskip
\begin{tabular}{cr@{.}lccccl}
\hline
\textbf{Name} & \multicolumn{2}{c}{$P_{\rm orb}$} & {\bf (SpT)}$_1$ & $M_2$ & $q$ & \textbf{Age} & \textbf{References, notes} \\
\hline
& \multicolumn{2}{c}{[d]} && [M$_\odot$] && [Myr] & \\
\hline
$\nu$~Cen & 2&6253 & B2\,IV & 0.91$_{-0.23}^{+0.44}$ & 0.13$^{+0.04}_{-0.06}$ & 11.1$_{-7.6}^{+5.2}$ & \cite{2021MNRAS.503.5554J} \\
$\gamma$~Lup A & 2&8498 & B2\,IV & 1.18$^{+0.75}_{-0.46}$ & 0.15$^{+0.09}_{-0.05}$ & 16.7$^{+5.0}_{-6.6}$ & \cite{2021MNRAS.503.5554J} \\
HD\,191495 & 3&6446 & B0\,V & 1.5$_{-0.3}^{+0.3}$ & 0.10$^{+0.03}_{-0.03}$ & 7.6$_{-1.8}^{+1.8}$ & Rotational variability?, \ \\ 
& \multicolumn{2}{c}{} &  & &&&  \cite{2023MNRAS.525.1641N} \\
V390\,Pup & 3&9277 & B1.5\,III & 1.27$^{+0.17}_{-0.19}$ & 0.14$^{+0.07}_{-0.04}$ & 16.2$_{-1.0}^{+1.0}$ & SPB/$\beta$~Cep + stochastic  \\
& \multicolumn{2}{c}{} &  & &&& variability?, this work, \\
& \multicolumn{2}{c}{} &  & &&& \cite{1978A+AS...32..409G} \\
HD\,149834 & 4&5957 & B2\,V & 0.86$_{-0.10}^{+0.13}$ & 0.087$_{-0.008}^{+0.010}$ & 5.9$_{-1.8}^{+3.9}$ & SPB/$\beta$~Cep + rotational\\
& \multicolumn{2}{c}{} &  & &&&  + stochastic, \cite{2021ApJ...910..133S},  \\
& \multicolumn{2}{c}{} &  & &&&  \cite{2024AJ....167...12C} \\
HD\,25631 & 5&2404 & B3\,V & 1.0$_{-0.2}^{+0.2}$ & 0.15$^{+0.04}_{-0.04}$ & 9.6$_{-9.4}^{+9.4}$ & SPB/$\beta$~Cep variability + TEOs?,\\ 
& \multicolumn{2}{c}{} &  & &&&  \cite{2023MNRAS.525.1641N} \\
c$^2$\,Sco & 5&7710 & B2.5\,Vn & 1.1$^{+0.07}_{-0.08}$ & 0.15$_{-0.04}^{+0.05}$ & 8$^{+6}_{-3}$ & SPB/$\beta$~Cep? variability, this \\
& \multicolumn{2}{c}{} &  & &&&  work, \cite{2016AJ....152...40G} \\
$\lambda$~Sco & 5&9525 & B2\,IV & 1.8$_{-0.2}^{+0.2}$ & 0.16$_{-0.03}^{+0.04}$& 11.5$_{-1.5}^{+1.5}$ & $\beta$~Cep variability, \\
& \multicolumn{2}{c}{} &  & &&&  \cite{2004A+A...427..581U}, \\
& \multicolumn{2}{c}{} &  & &&&  \cite{2006MNRAS.370..884T}, triple star \\
HD\,46485 & 6&9431 & O7\,V & 0.95$_{-0.15}^{+0.15}$ & 0.041$^{+0.007}_{-0.010}$ & 2.5$_{-1.1}^{+1.1}$ & Rotational + stochastic  \\ 
& \multicolumn{2}{c}{} &  & &&&  variability?, \cite{2023MNRAS.525.1641N} \\
$\mu$~Eri & 7&3806 & B5\,IV & 0.77$_{-0.02}^{+0.02}$ & 0.124$_{-0.007}^{+0.008}$ & 62$_{-7}^{+7}$ & SPB variability, \\
& \multicolumn{2}{c}{} &  & &&& \cite{2020ApJ...903...96G}, \\
& \multicolumn{2}{c}{} &  & &&& \cite{2013MNRAS.432.1032J} \\
16 (EN) Lac & 12&0968 & B2\,IV & 1.29$_{-0.07}^{+0.07}$ & 0.13$^{+0.03}_{-0.02}$ & 16.3$_{-1.5}^{+1.5}$ & $\beta$~Cep variability,\\
& \multicolumn{2}{c}{} &  & &&&  \cite{2015MNRAS.454..724J} \\
& \multicolumn{2}{c}{} &  & &&&  \cite{1988AcA....38..401P} \\
\hline
\end{tabular}
\end{table}

Following on from the Introduction, it seems that the most effective method of searching for new Galactic NBs is to survey the light curves of eclipsing binaries with a massive primary component, that is, with an O or early B-type primary. Systems with strong irradiation effect will be good candidates for NBs. In this context, it seems that TESS observations should be a particularly good source of data. 
%\comm{[+ In addition it is necessary to have the multi-epoch RV observations! (see above)]}  
With reference to Fig.~8, in order to verify the evolutionary status of the secondary components, it is also important to determine the age of the system. For such determinations, comparison of the primary component's parameters with models is most commonly used. 
%\comm{We are dealing with OB stars, and most of them have already been interacting, and if the primary got rejuvenated, and we don't know without RV curve is the post- or pre-interacting system..? Just be a bit careful, that first we need to prove that it's the pre-interacting system with a pre-MS companion..} 
An alternative is the determination of the age of the parent open cluster or the OB association, provided that the system is a member.

\begin{acknowledgments}
This work was supported by the National Science Centre, Poland, grant no.~2022/45/B/ST9/ 03862. We thank Pascal Torres from PUC for providing us with the preliminary radial velocities of V390~Pup derived from the FIDEOS spectra. Based on data collected by the BRITE Constellation satellite mission, designed, built, launched, operated and supported by the Austrian Research Promotion Agency (FFG), the University of Vienna, the Technical University of Graz, the University of Innsbruck, the Canadian Space Agency (CSA), the University of Toronto Institute for Aerospace Studies (UTIAS), the Foundation for Polish Science \& Technology (FNiTP MNiSW), and National Science Centre (NCN). This work used the SIMBAD and Vizier services operated by Centre des Donn\'ees astronomiques de Strasbourg (France), and bibliographic references from the Astrophysics Data System maintained by SAO/NASA. This paper includes data collected by the TESS mission. Funding for the TESS mission is provided by the NASA's Science Mission Directorate.
\end{acknowledgments}

\begin{furtherinformation}

\begin{orcids}
\orcid{0000-0003-2488-6726}{Andrzej}{Pigulski}
\end{orcids}

\begin{conflictsofinterest}
The author declares no conflict of interest.
\end{conflictsofinterest}

\end{furtherinformation}

\bibliographystyle{bullsrsl-en}

\bibliography{BRITE-Nascent}

\end{document}